# Flexible Full-Stokes Polarization Engineering by Disorder-Scrambled Metasurfaces

*Zhi Cheng, Zhou Zhou, Zhuo Wang, Yue Wang, Changyuan Yu \**

*Z. Cheng, Y. Wang, C. Yu*

*Department of Electrical and Electronic Engineering,*

*the Hong Kong Polytechnic University, Hong Kong SAR, China*

*changyuan.yu@polyu.edu.hk*

*Z. Zhou*

*Department of Electrical and Computer Engineering, and NUS Graduate School,*

*National University of Singapore, Singapore 117583, Singapore*

*Z. Wang*

*Guangdong Provincial Key Laboratory of Nanophotonic Functional Materials and*

*Devices, School of Optoelectronic Science and Engineering,*

*South China Normal University 510006, Guangzhou, China*

*C. Yu*

*The Hong Kong Polytechnic University Shenzhen Research Institute, Shenzhen, China*

**Abstract:** The ability to arbitrarily and flexibly control the polarization of light, including both the state of polarization (SoP) and the degree of polarization (DoP), is highly important for quantum optics, polarization imaging, and coherent optical communications. Although metasurfaces have shown promise in polarization control, the few studies focusing on the DoP often lack flexibility in manipulation. Here, we propose a novel approach using a disordered metasurface to flexibly convert natural light into partially polarized light, enabling independent and flexible control over all Stokes parameters. The metasurface is composed of two types of meta-atoms, uniformly distributed with specific quantity ratios, decoupling the design parameters in the process of polarization control, and allowing a one-to-one correspondence between metasurface and polarization spaces. The azimuthal and elevation angles of the SoP on the Poincaré sphere are independently controlled by the meta-atom rotation and size, while the DoP is governed by the quantity ratio. A developed algorithm determines the disordered metasurface arrangement, with theoretical calculations showing an average error of less than 3° for both the azimuthal and elevation angles and a control accuracy of ±0.05 for the DoP.

# 1. Introduction

The capability to precisely manipulate the full-stokes parameters of light, encompassing both the state of polarization (SoP) and degree of polarization (DoP), is a cornerstone of photonics research and development. In deterministic systems, polarization refers to the orientation of the electric field vector, which defines the SoP. In stochastic systems, polarization evolves over time, giving rise to light that may be unpolarized or partially polarized. This variability is quantitatively described by the DoP, which measures the fraction of light that remains polarized. Partially polarized light is thus a combination of fully polarized (dominate polarization) and naturally unpolarized components, with the DoP indicating the relative contribution of the polarized component.

Controlling SoP and DoP is vital for applications such as optical communication [1-3], quantum optics [4-7], imaging [8, 9], remote sensing [10], and spectroscopy [11-13]. SoP manipulation is commonly achieved through the use of wave plates, which introduce controlled phase shifts between orthogonal polarization components. On the other hand, DoP modulation typically involves converting fully polarized or unpolarized light into a desired partially polarized state. Achieving this requires differential transmission between orthogonal polarization states, as demonstrated in Figure 1(a). This is often implemented experimentally using time-varying optical retarders or wavelength-dependent wave plates. Simultaneous control over both SoP and DoP, however, presents significant engineering challenges. Traditional approaches for such control often rely on complex and bulky optical setups, which can impede the integration and miniaturization of photonic devices. These limitations highlight the need for more compact and efficient solutions capable of fully manipulating the polarization properties of light.

Recently, metasurfaces—comprising arrays of nanoscale structures—have garnered significant attention as flat optical devices capable of manipulating light's amplitude, phase, and polarization at subwavelength scales. These capabilities have catalyzed advancements in fields such as imaging [14-19], polarization control [20-34], quantum optics [35-39], and communications [40-45]. Importantly, metasurfaces facilitate the miniaturization and integration of photonic devices, offering enhanced functionality within compact platforms. As shown in Figure 1(a), in the context of polarization control, a mapping relationship between the metasurface design parameter space and the polarization control parameter space can be established through careful design. The birefringent meta-atoms, shown in Figure 1(b), function as a nano-waveplate widely used in polarization modulation [46-48]. When the transmittance is set close to unity, the system becomes effectively lossless, yielding a unitary Jones matrix. This allows the metasurface to convert one SoP into another by redistributing phase delay between orthogonal polarization components. However, in dynamic systems with incoherent light, time-averaging equalizes intensity differences between these components, rendering this approach unsuitable for controlling the DoP. By carefully adjusting the gap distance between two types of meta-atoms, —within the "weak coupling" region [20, 31]—far-field radiation interference is induced in the diatomic structure shown in Figure 1(c). This design could provide at least six degrees of freedom, enabling the realization of any arbitrary Jones matrix for polarization conversion [22, 23, 26, 28, 29, 31]. The flexibility of this method allows for the creation of non-unitary Jones matrices, which generate intensity differences between orthogonal states in a time-varying system. However, this results in an intrinsic coupling between the metasurface design parameters, specifically the dominant SoP and the DoP within the polarization space, as shown in Figure 1(c). This relationship can be further understood using the phasor diagram presented in Figure S1. Wang et al. recently tackled this challenge by introducing an inverse-designed meta-atom, optimized through topological methods, capable of converting unpolarized light into partially polarized light [32]. However, the complex geometries of these optimized structures require high-precision fabrication, which presents significant challenges for practical implementation.

In this study, we propose a novel method to decouple the SoP and the DoP using a disordered metasurface. The disordered metasurface consists of two types of birefringent meta-atoms uniformly distributed across the entire surface, as shown in Figure 1(d). By leveraging far-field interference effects among these meta-atoms, we introduce polarization-dependent loss (PDL), enabling precise modulation of the DoP. The DoP is determined by the differential conversion of orthogonal polarization components, which, in turn, is flexibly governed by the quantity ratio of different types of meta-atoms within the disordered metasurface. Control over the SoP is achieved by independently adjusting the azimuth and elevation angles on the Poincaré sphere through the rotation and the phase difference of individual meta-atoms. To generate the disordered metasurface, we developed an algorithm based on a greedy search combined with a 2D bin-packing problem. In this design, the phase differences, rotation angles, and quantity ratios of meta-atoms collectively determine the polarization both on and within the Poincaré sphere, enabling access to all Stokes parameters. We theoretically demonstrate that the evolution of polarization states on the Poincaré sphere can be systematically controlled by varying a single parameter, producing polarization distributions along the latitude, longitude, and radial directions. The error in controlling the DoP is approximately ±0.05, while the average errors in the azimuth and elevation angles of the dominant SoP are both less than 3°. We further demonstrate that this design layout possesses a certain degree of tolerance, as different meta-atom combinations can function effectively within the same framework. Additionally, we demonstrate the flexibility of this design approach in enabling arbitrary polarization state conversion applications.

## 2. Results and discussion

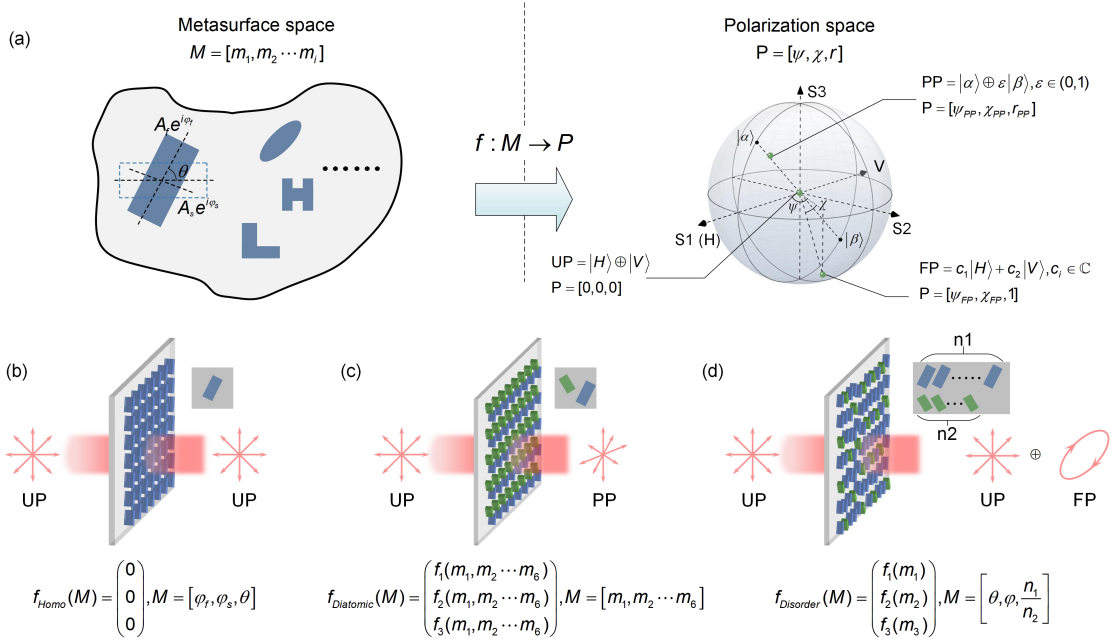

Figure 1 (a) The mapping between metasurface parameter space and polarization control space. The metasurface parameters include the complex transmission coefficients of the fast and slow axes, as well as the rotation angle. A point on or inside the Poincaré sphere is defined by the azimuth $\psi$, elevation angles $\chi$, and radius $r$. The symbol $\oplus$ represents the incoherent superposition of light. PP: partially polarized, UP: unpolarized, FP: fully polarized. (b) The homogeneous metasurface with one meta-atom in a unit cell, cannot convert unpolarized light into partially polarized light. (c) The diatomic metasurface with two meta-atoms in a supercell, converts unpolarized light into partially polarized light. (d) The proposed disordered metasurface could partially polarize natural light with the independent control of the dominant full-polarized light and DoP.

The physical mechanisms that simultaneously and independently control DoP and SoP using metasurfaces are illustrated in Figure 2. Since unpolarized light can be viewed as an incoherent superposition of two orthogonal polarization states, analyzing this mechanism requires focusing on the conversion process between these orthogonal states. Each metasurface is capable of converting ($|\alpha^*\rangle$, $|\beta^*\rangle$) into ($|\alpha\rangle$, $|\beta\rangle$), with distinct phase shifts for the respective states. This results in constructive interference for the $|\alpha\rangle$ state and destructive interference for the $|\beta\rangle$ state. When the two metasurfaces differ in their proportional contribution, the degrees of constructive interference for $|\alpha\rangle$ and destructive interference for $|\beta\rangle$ are no longer balanced, leading to an unequal ratio of $|\alpha\rangle$ and $|\beta\rangle$ in the final incoherent light output. In essence, differential conversion between $|\alpha\rangle$ and $|\beta\rangle$ is achieved, resulting in partially polarizing the unpolarized light. In what follows, the derivation of the necessary Jones matrices, the mechanism for controlling DoP, and the design methodology for the metasurfaces are discussed.

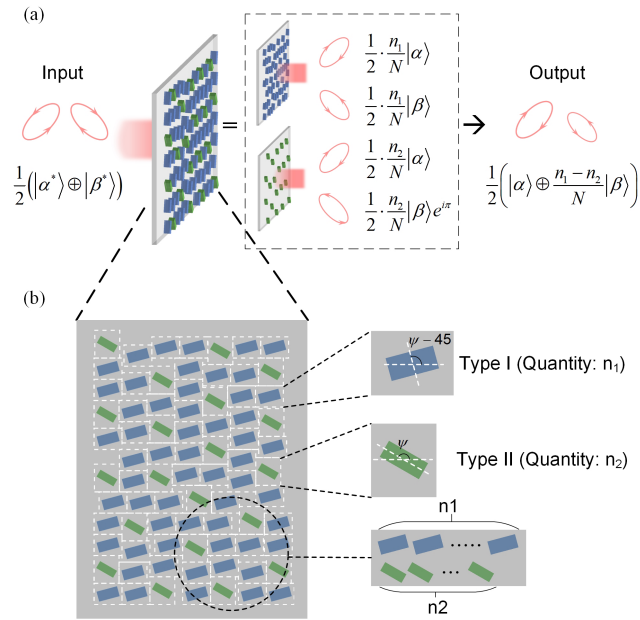

Figure 2 (a) The physical mechanism of partially polarized natural light with the independent control of the SoP and DoP using metasurfaces. (b) The construction of the disordered metasurface. The green and blue rectangles represent the top views of different meta-atoms. The white dashed box represents the effective range of each meta-atom.

We begin with the birefringent meta-atoms shown in Figure 1(a) which are commonly used for polarization control. The behavior of these meta-atoms is described by the following Jones matrix:

$$J_{atom} = R(\theta)\begin{pmatrix} A_f e^{i\varphi_f} & 0 \\ 0 & A_s e^{i\varphi_s} \end{pmatrix} R(-\theta) \quad (1)$$

This Jones matrix accounts for the phase retardation along the fast and slow optical axes, denoted by $\varphi_f$ and $\varphi_s$, respectively, with $A_f$ and $A_s$ representing the transmittances along these axes. The rotation matrix $R(\theta)$, defined $R(\theta) = \begin{bmatrix} \cos(\theta) & -\sin(\theta) \\ \sin(\theta) & \cos(\theta) \end{bmatrix}$, describes the orientation of the birefringent axes with respect to the incident polarization. With transmittances close to unity, the system becomes effectively lossless, yielding a unitary Jones matrix. To introduce differential transmission between orthogonal states, a non-unitary Jones matrix is required, which can be constructed as a sum of unitary matrices.

On the other hand, the Jones vectors of an arbitrary orthogonal polarization pair can be expressed in terms of azimuth angle $\psi$ and elevation angle $\chi$ on the Poincare sphere as:

$$\alpha = R(\psi - 45°)\begin{pmatrix} e^{-i\chi} \\ e^{i\chi} \end{pmatrix}, \beta = R(\psi - 45°)\begin{pmatrix} e^{-i\chi} \\ -e^{i\chi} \end{pmatrix} \quad (2)$$

Here, $|\alpha\rangle$ is located at coordinates $(2\psi, 2\chi)$ on the Poincaré sphere, while $|\beta\rangle$ is positioned at $(2\psi - 180°, -2\chi)$, symmetrically opposite about the origin. It is important to note that, in the polarization states $(|\alpha\rangle, |\beta\rangle)$, the x- and y-components are complex conjugates, except for the negative sign in $|\beta\rangle$. This indicates that by applying a phase delay to the x- and y-components, as described by the Jones matrix in the form of Equation (1), it is possible to convert between the polarization states $(|\alpha^*\rangle, |\beta^*\rangle)$ and their conjugate states $(|\alpha^*\rangle, |\beta^*\rangle)$ (Supplemental Information Note II). Therefore, the Jones matrix for converting the orthogonal conjugate states can be expressed in a form similar to Equation (1):

$$J_1 = R(\psi - 45°)\begin{pmatrix} e^{-2i\chi} & 0 \\ 0 & e^{2i\chi} \end{pmatrix} R(45° - \psi) \quad (3)$$

This Jones matrix converts the conjugate orthogonal states ($|\alpha^*\rangle$, $|\beta^*\rangle$) into ($|\alpha\rangle$, $|\beta\rangle$) while introducing the same global phase. This means that there is no relative phase shift between the two states. Another Jones matrix can be constructed to achieve the conversion between these two pairs of conjugate orthogonal states with a relative phase difference of π (see Supplemental Information Note II). The Jones matrix for this transformation can be expressed as:

$$J_2 = R(\psi)\begin{pmatrix}1 & 0\\ 0 & -1\end{pmatrix}R(-\psi) = R(\psi - 45)\begin{pmatrix}0 & 1\\ 1 & 0\end{pmatrix}R(45 - \psi) \quad (4)$$

This matrix flips the x- and y-components of the Jones vector in Equation (2). The negative sign in the y-component of the $|\beta\rangle$ state introduces a global phase of π during the conjugate state conversion process. By combining the Jones matrices from Equation (3) and (4) with different coefficients, differential transmission between orthogonal polarization states can be achieved, as shown in Figure 2(a). The resulting composite Jones matrix can be expressed as follows:

$$J = J_1 + J_2 = \frac{n_1}{N}R(\psi - 45°)\begin{pmatrix}e^{-2i\chi} & 0\\ 0 & e^{2i\chi}\end{pmatrix}R(45° - \psi) + \frac{n_2}{N}R(\psi)\begin{pmatrix}1 & 0\\ 0 & -1\end{pmatrix}R(-\psi)$$
$$(5)$$

To ensure energy conservation, the sum of the coefficients must be equal to unity. Therefore, the coefficients are written in the form of $n_i/N$, where i = 1, 2, and $N = n_1 + n_2$. The state $|\alpha^*\rangle$ is fully converted into $|\alpha\rangle$ with an amplitude of 1/2. However, the conversion between $|\beta^*\rangle$ and $|\beta\rangle$ undergoes destructive interference due to the π-phase delay, resulting in an amplitude of $(n_1 - n_2)/2N$. Consequently, the transmission matrix for converting ($|\alpha^*\rangle$, $|\beta^*\rangle$) into ($|\alpha\rangle$, $|\beta\rangle$) is given by:

$$J_t = \begin{pmatrix}t_{\alpha^*\alpha} & t_{\beta^*\alpha}\\ t_{\alpha^*\beta} & t_{\beta^*\beta}\end{pmatrix} = \begin{pmatrix}1 & 0\\ 0 & \frac{n_1-n_2}{N}\end{pmatrix} \quad (6)$$

To determine the DoP in the time domain, it is necessary to apply a time-averaged operation to account for the random phase in unpolarized light. The intensity of the outgoing light can then be expressed as:

$$\langle I_{out} \rangle = (1 - t_{\beta^*\beta}^2)\langle \alpha \rangle + t_{\beta^*\beta}^2(\langle \alpha \rangle + \langle \beta \rangle)$$
$$= (1 - t_{\beta^*\beta}^2)FP + 2t_{\beta^*\beta}^2 UP \quad (7)$$

where the brackets <> denote to time average. According to the definition of DoP, let $\eta = n_1/n_2$, it is calculated as:

$$p = \frac{1-t_{\beta^*\beta}^2}{1+t_{\beta^*\beta}^2} = \frac{2\eta}{1+\eta^2} \quad (8)$$

To achieve the Jones matrix in Equation (5), we propose a disordered metasurface composed of various types of meta-atoms. These adjacent meta-atoms are arranged with appropriate gap distances in the weak coupling regime. In this case, the Jones matrix is the summation of the matrices of each meta-atom, represented in the following form:

$$J_{DM} = \sum_{i=1}^{M} \frac{n_i}{N} J_{atom}^i \quad (9)$$

where $n_i$ ($i=1,2…M$) represents the number of each meta-atom, and $N$ denotes the total number of atoms. When $M=2$ and $n_1 = n_2$, the disordered metasurface degenerates into the simpler diatomic design shown in Figure 1(c), where a periodic unit cell within wavelength scale is constructed across the metasurface. However, when $n_1 \neq n_2$, if we continue to follow the conventional periodic design approach, multiple types of meta-atoms must be arranged within a supercell, resulting in a lattice constant larger than the wavelength. This oversized lattice constant can introduce energy losses by diverting energy away from the intended polarization conversion, leading to the generation of higher-order diffraction modes. Additionally, while a periodic design does not inherently produce a phase gradient, it is more prone to introducing one without careful optimization. In the presence of a phase gradient, the periodicity can amplify this effect, leading to stronger higher-order diffraction and further deflection of the outgoing beam as shown in Figure S2. More critically, the periodic design complicates the ability to maintain a uniform distribution of meta-atoms across the metasurface. The translational symmetry inherent to periodic structures hinders the preservation of a consistent ratio of meta-atoms near the boundaries, thereby diminishing polarization conversion efficiency (see Supplemental Information Note III).

Building on this understanding, the disordered metasurface, as illustrated in Figure 2(b), consists of $n_1$ meta-atoms of type I and $n_2$ meta-atoms of type II. Each type of meta-atom is uniformly arranged, with their spacing optimized for the weak-coupling region to ensure effective interaction. This design ensures far-field radiation coherence among the meta-atoms, which is essential for the validity of Equation (9), where the total response of the metasurface is expressed as the summation of the individual meta-atoms Jones matrices. If the meta-atoms are placed too close together, strong coupling occurs, altering the propagation phase. Conversely, if they are too far apart, the meta-atoms interact weakly, and no coherent interference is observed. The uniform arrangement ensures that the local ratio of meta-atoms within each effective coupling region remains consistent with the target ratio $n_1/n_2$ across the entire surface. Unlike conventional interleaved designs, where different meta-atom types function independently, the meta-atoms in the disordered metasurface work cooperatively. The intentional disorder breaks the translational symmetry, effectively suppressing strong higher-order diffraction modes. Through this design, the Stokes parameters (S1, S2, S3) and the DoP are independently controlled. The polarization orientation, represented by S1, S2, and S3, can be mapped to the azimuth and elevation angles on the Poincaré sphere, which is governed by the rotation angle and phase delay of the meta-atoms, as described by Equation (5). Meanwhile, the DoP, corresponding to the radius in the Poincaré sphere, is controlled by the ratio of the two types of meta-atoms ($n_1/n_2$) within the disordered metasurface. This decoupled control allows for independent tuning of the polarization state's orientation and its purity, enabling full manipulation of all relevant Stokes parameters.

To obtain the desired arrangement of meta-atoms, we developed a two-dimensional bin-packing algorithm based on a greedy search heuristic (see Supplemental Information Note IV). Our empirical calculations show that the weak coupling region typically spans about 1/10 of the operating wavelength. We define the effective working region of each meta-atom using the enclosing rectangles, as shown by the dashed lines in Figure 2(b). These rectangles are tessellated uniformly across the metasurface by the algorithm, ensuring that the arrangement maintains both uniformity and the adaptive spacing required for coherent interactions. While our approach yields strong results, other advanced algorithms could also be employed to achieve similarly optimized layouts for disordered metasurfaces.

We validate the proposed concept by demonstrating full control of the SoP and DoP across the entire Poincaré sphere through simulations conducted using the Finite-Difference Time-Domain (FDTD) method, as shown in Figure 3. For fully polarized light, the polarization state is represented by points on the surface of the Poincaré sphere, while partially polarized light is depicted by points within the sphere. Complete polarization control requires the manipulation of three parameters on the Poincaré sphere: azimuthal angle, elevation angle, and radius, which correspond to the Stokes parameters S1, S2, S3, and the DoP. By independently varying these parameters, we demonstrate precise control over the latitude, longitude, and radius of the Poincaré sphere, allowing for arbitrary manipulation of partially polarized light.

In the simulation, 500 points are uniformly sampled across the surface of the Poincaré sphere, with near-equal Euclidean distances between adjacent points, to represent the input of unpolarized light. The random phase component inherent in unpolarized light is neglected, as it does not influence the results in time-averaged incoherent superposition calculations. The polarization states corresponding to the sampled points are converted by the designed metasurface, resulting in a non-uniform distribution of Stokes vectors across both the surface and interior of the Poincaré sphere, as shown in Figure 4(a). The incoherent sum of these transformed polarization states yields the DoP and the dominant SoP.

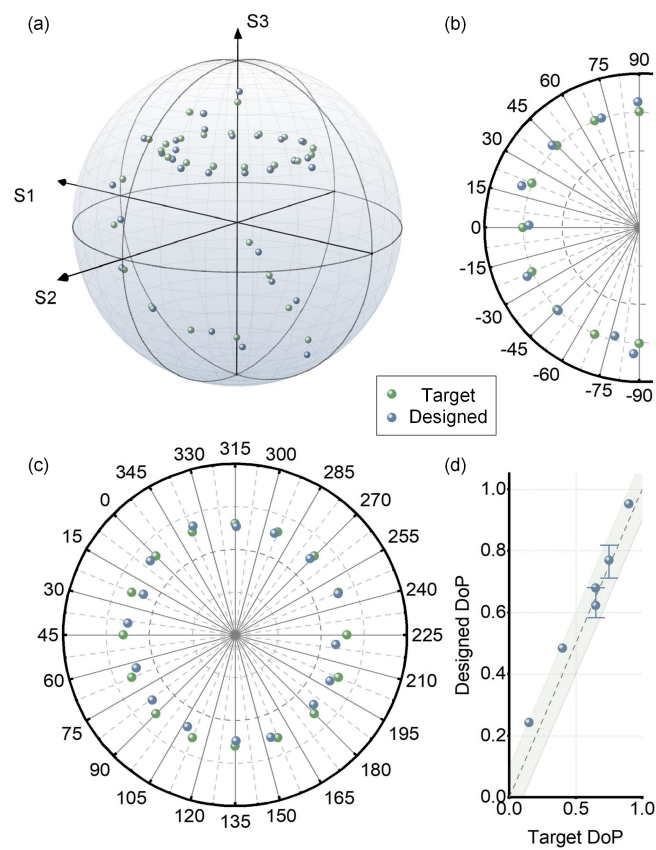

Figure 3 Demonstration of independent control of the azimuth and elevation angles of the dominant polarization state and the DoP. (a) Evolution of the polarization on the Poincare sphere along the latitude at $2\chi = 50°$ with DoP=0.65, longitude at $2\psi = 45°$ with DoP = 0.75, and radius at $2\chi = -60°$, $2\psi=225°$, respectively. The green dots represent the target polarization while the blue dots represent the calculated results. (b) The side view of (a) illustrates the polarization evolution along the longitude. (c) The top view of (a) illustrates the polarization evolution along the latitude. (d) The DoP values of all calculation points. The scatter points are along the radius, and the other two error bars are along the longitude and latitude respectively. The green shaded area represents the range of DoP error ±0.1.

Figure 3 illustrates the latitude and longitude lines on the Poincaré sphere passing through the polarization state at $2\chi = 50°$ and $2\psi = 45°$. These polarization states are achieved by varying the size or orientation of two types of meta-atoms. Only half of the longitude is depicted, as the other half can be obtained by rotation of the meta-atoms. To enhance visualization clarity, the dominant SoP is chosen as ($2\chi = -60°$, $2\psi=225°$), located in the southern hemisphere, when demonstrating DoP control through adjustments to the meta-atom quantity ratio. For the latitude data set, the DoP is set to 0.65, while for the longitude data set, it is set to 0.75, corresponding to quantity ratios of 2.714 and 2.214, respectively. In the metasurface design, the target DoP values are approximated by quantity ratios of 19:7 and 31:14 for the two types of meta-atoms. The arrangement coordinates and structural parameters of the employed meta-atoms are detailed in Supplemental Information Note V. In Figure 3, the green points represent the target polarization states on the Poincaré sphere, while the blue points represent the simulation results. The longitude and latitude evolution data sets are also shown in Figure 3(b) and Figure 3(c), with side-view and top-view perspectives, respectively. The average error in the elevation angle is 2.65°, while the average error in the azimuth angle is 2.98°, demonstrating the accuracy of this method in controlling the polarization state. In Figure 3(d), error bars indicate the range of simulated DoP values for the latitude and longitude data sets, with the central sphere symbol representing the average DoP. The green shaded area corresponds to a DoP error range of ±0.1, while nearly all data points fall within the tighter range of ±0.05. As the elevation angle increases, we observe a rise in error. This is attributed to the limited size of the meta-atom database (see Supplemental Information Note V) and the relatively large size difference between the two types of meta-atoms in the high-latitude regions. This discrepancy complicates the task of finding an appropriate spacing to satisfy both weak coupling and coherence conditions. The mismatch between the propagation phase of the selected meta-atoms and the target phase could also contribute to the observed errors. Expanding the meta-atom database could help mitigate both issues and reduce the error.

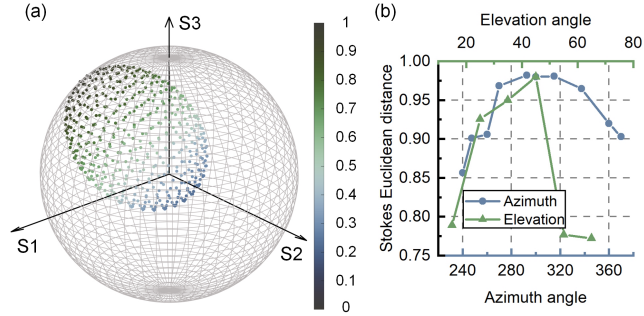

Figure 4 (a) Distribution of all polarization states of partially polarized light ($2\psi=300°$, $2\chi=45°$, DoP=0.65) in the time domain on a solid Poincare sphere. The color represents the light intensity and is normalized. (b) Robustness of disordered metasurface design. Taking the position of partially polarized light represented by (a) on the Poincare sphere as the origin, the disordered metasurface is constructed using the same layout along the longitude and latitude directions. The Stokes distance is used to represent the accuracy of the polarization state conversion. The intersection of the two lines is the state represented by (a).

To further demonstrate the feasibility and flexibility of this approach, we showcase the ability to generate different partially polarized light using the same metasurface arrangement. Since the coherent pixels operate in a weak coupling regime, small variations in the gap distance between adjacent meta-atoms have a negligible effect on the modulation behavior. This introduces a tolerance in the gap distance, allowing the same metasurface design to be applied across different modulation scenarios.

For instance, the layout designed to achieve a DoP of 0.65, with a quantity ratio of 2.714, is partially illustrated in Figure 2(b). The full arrangement parameters are listed in Supplemental Information Note V. In this layout, the effective sizes of the meta-atoms are (850, 650) nm, and (760, 850) nm. The dominant polarization state is initialized at $2\psi = 300°$ and $2\chi = 45°$, as shown in Figure 4(a). The accuracy of the partially polarized light conversion is quantified by the Stokes Euclidean Distance (SED), which is defined as 1−distance, where "distance" represents the Euclidean distance between the measured and target Stokes vectors. A value of SED = 1 indicates perfect alignment (i.e., zero Euclidean distance) between the two Stokes vectors.

Figure 4(b) also illustrates the operational range of the disordered metasurface, defined as the effective polarization conversion range in terms of latitude and longitude on the Poincaré sphere. Starting from the initial point, we varied the sizes and rotation angles of the meta-atoms, causing the partially polarized light to move along both the longitude and latitude lines passing through the initial point. As shown in Figure 4(b), the intersection of these two lines represents the initial point. The tolerance in the azimuthal angle (meta-atom rotation) is approximately 60° (with the range of SED exceeding 0.95), while the elevation angle (meta-atom size) tolerance is less than 10°. This discrepancy arises because adjusting the elevation angle requires a complete change in the meta-atom combination, for which the initial effective sizes are not suitable. Although the physical gap between different meta-atom combinations may exhibit varying tolerances, the specific values presented here are not universal. Nonetheless, these results confirm that the metasurface design provides a tolerance margin, allowing the same layout to be used for different meta-atom configurations. This tolerance is crucial for practical applications, as it enables flexible and independent control of both the DoP and SoP.

## 3. Conclusion

In conclusion, we propose a disordered metasurface capable of converting unpolarized light into partially polarized light, with independent control over both the SoP and DoP. Our approach leverages far-field interference between meta-atoms and introduces the concept of a quantity ratio between different types of meta-atoms, enabling arbitrary and flexible control over polarization states. Furthermore, by generalizing this approach, we provide an analytical solution where each element of an arbitrary Jones matrix can be directly expressed using the optical parameters of the meta-atoms, namely their propagation phase and orientation, as detailed in the Supplemental Information Note VI using a triatomic design. In contrast to inverse design methods, our approach provides a more intuitive understanding of the underlying physics. It establishes a concise and direct relationship between the meta-atoms optical parameters, the Jones matrix, and full Stokes parameter control. Theoretically, this approach introduces a new degree of freedom in metasurface design—namely, the ratio of quantities—and allows for the synergistic effect of multiple meta-atoms. The expanded DoFs suggest that the method could be extended to polarization-dependent applications that are decoupled from wavelength constraints, such as multi-color or broadband systems, or multichannel polarization multiplexing applications. This method enables flexible control over multi-dimensional light fields, positioning it as a promising tool for advanced applications in imaging, communication, and sensing.

**Acknowledgements**

We thank Prof. Cheng-wei Qiu for valuable comments and feedback. The authors thank the support of Hong Kong Research Grants Council (GRF 15209321 B-Q85G).

Supplemental Information for

# Flexible Full-Stokes Polarization Engineering by Disorder-Scrambled Metasurfaces

*Zhi Cheng, Zhou Zhou, Zhuo Wang, Yue Wang, Changyuan Yu \**

*Z. Cheng, Y. Wang, C. Yu*

*Department of Electrical and Electronic Engineering,*

*the Hong Kong Polytechnic University, Hong Kong SAR, China*

*changyuan.yu@polyu.edu.hk*

*Z. Zhou*

*Department of Electrical and Computer Engineering, and NUS Graduate School,*

*National University of Singapore, Singapore 117583, Singapore*

*Z. Wang*

*Guangdong Provincial Key Laboratory of Nanophotonic Functional Materials and Devices, School of Optoelectronic Science and Engineering,*

*South China Normal University 510006, Guangzhou, China*

*C. Yu*

*The Hong Kong Polytechnic University Shenzhen Research Institute, Shenzhen, China*

**Note I: Coupling Between Dominant State of Polarization and Degree of Polarization in diatomic design**

Under the weak-coupling regime, the Jones matrix of the diatomic metasurface can be expressed as the sum of each meta-atom's Jones matrix. This means that arbitrary Jones matrices can be constructed using this approach. The summation process of the Jones matrix elements is illustrated in Fig. S1: the small blue dashed circle represents the unit vector, indicating the complex amplitude of the combined Jones matrix elements. The large blue region, with a modulus of 2, represents the total magnitude of the combined vectors.

The initial combined vector, shown as the green arrow with an amplitude of A1, can be decomposed into two unit vectors. When attempting to adjust A1 to A1'—that is, when trying to change the polarization conversion properties—simply adjusting the surface parameters results in a new combined vector A1' (yellow arrow). However, we observe that the new unit vectors differ substantially from the original ones. This indicates that in the design of diatomic metasurfaces for polarization control, adjustments to polarization conversion properties affect all variables, and modifying just one or a few parameters will not suffice to achieve the desired results.

In other words, practical design requires finding new combinations of meta-atoms and adjusting rotation angles and other parameters to meet the design requirements.

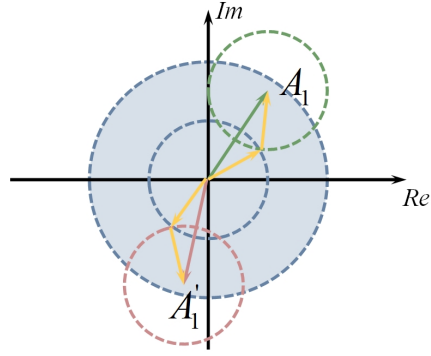

Fig. S1. Phasor diagram of the summation process of Jones matrix elements.

**Note II: Derivation of the Jones Matrix for Independent Control of SoP and DoP**

Given an arbitrary elliptical polarization state, the states α and β are expressed as:

$$\alpha = R(\psi - 45°)\begin{pmatrix} e^{-i\chi} \\ e^{i\chi} \end{pmatrix}, \beta = R(\psi - 45°)\begin{pmatrix} e^{-i\chi} \\ -e^{i\chi} \end{pmatrix} \quad (S1)$$

The corresponding conjugate polarization states, **α\*** and **β\***, are represented as：

$$\alpha^* = R(\psi - 45°)\begin{pmatrix} e^{i\chi} \\ e^{-i\chi} \end{pmatrix}, \beta^* = R(\psi - 45°)\begin{pmatrix} e^{i\chi} \\ -e^{-i\chi} \end{pmatrix} \quad (S2)$$

We aim to find a Jones matrixes J that either convert (**α\***, **β\***) into (**α, β**) or (**α, βe^{iπ}**). Thus, we seek：

$$\begin{aligned} J_1(\alpha^*, \beta^*) &= (\alpha, \beta) \\ J_2(\alpha^*, \beta^*) &= (\alpha, \beta e^{-i\pi}) \end{aligned} \quad (S3)$$

Noting that the Jones matrix for this system, as shown in previous works, must involve a rotation matrix combined with a diagonal matrix, we first rotate the Jones matrix in alignment with the elliptical polarization states, which corresponds to a rotation by $\psi - 45$:

$$J = R(\psi - 45°)DR^{-1}(\psi - 45°) \quad (S4)$$

where the matrix D is of the form:

$$D = \begin{pmatrix} d_{11} & 0 \\ 0 & d_{22} \end{pmatrix} \quad (S5)$$

Substituting Equation (S4) into the first equation of (S3) yields:

$$J\alpha^* = R(\psi - 45°)DR^{-1}(\psi - 45°)R(\psi - 45°)\begin{pmatrix} e^{i\chi} \\ e^{-i\chi} \end{pmatrix} = R(\psi - 45°)D\begin{pmatrix} e^{i\chi} \\ e^{-i\chi} \end{pmatrix} = \alpha \quad (S6)$$

It is found that for J1, the same equation is obtained for both α and β. Since R is an invertible matrix, the above equation can be simplified to:

$$D\begin{pmatrix} e^{i\chi} \\ e^{-i\chi} \end{pmatrix} = \begin{pmatrix} e^{-i\chi} \\ e^{i\chi} \end{pmatrix} \quad (S7)$$

Solving the equations, we get:

$$\begin{aligned} d_{11}e^{i\chi} = e^{-i\chi} &\Rightarrow d_{11} = e^{-2i\chi} \\ d_{22}e^{-i\chi} = e^{i\chi} &\Rightarrow d_{22} = e^{2i\chi} \end{aligned} \quad (S8)$$

Thus, the required Jones matrix is:

$$J_1 = R(\psi - 45°)\begin{pmatrix} e^{-2i\chi} & 0 \\ 0 & e^{2i\chi} \end{pmatrix}R^{-1}(\psi - 45°) \quad (S9)$$

It is easy to see that in order to find a suitable J2, if the Jones matrix continues to rotate by ψ - 45°, substituting Equation 4 into the second term of Equation 3 yields no solution. Therefore, different rotation angles need to be considered by solving the following system of equations:

$$\begin{aligned} DR(\theta)\begin{pmatrix} e^{i\chi} \\ e^{-i\chi} \end{pmatrix} &= R(\theta)\begin{pmatrix} e^{-i\chi} \\ e^{i\chi} \end{pmatrix} \\ DR(\theta)\begin{pmatrix} e^{i\chi} \\ -e^{-i\chi} \end{pmatrix} &= R(\theta)\begin{pmatrix} -e^{-i\chi} \\ e^{i\chi} \end{pmatrix} \end{aligned} \quad (S10)$$

Where $\theta = \psi - 45° - \psi_2$, with $\psi_2$ being the rotation angle of J2. Expanding $R(\theta)$ gives the following two sets of equations:

$$\begin{cases} d_{11}\cos\theta \cdot e^{i\chi} - d_{11}\sin\theta \cdot e^{-i\chi} = e^{-i\chi}\cos\theta - e^{i\chi}\sin\theta \\ d_{22}\sin\theta \cdot e^{i\chi} + d_{22}\cos\theta \cdot e^{-i\chi} = e^{-i\chi}\sin\theta + e^{i\chi}\cos\theta \end{cases}$$
$$\begin{cases} d_{11}\cos\theta \cdot e^{i\chi} + d_{11}\sin\theta \cdot e^{-i\chi} = -e^{-i\chi}\cos\theta - e^{i\chi}\sin\theta \\ d_{22}\sin\theta \cdot e^{i\chi} - d_{22}\cos\theta \cdot e^{-i\chi} = e^{-i\chi}\sin\theta + e^{i\chi}\cos\theta \end{cases} \quad (S11)$$

By comparing the exponents of the exponential terms on both sides of the equation, the following relationships can be derived:

$$\begin{cases} d_{11}\cos\theta = -\sin\theta \\ d_{11}\sin\theta = -\cos\theta \\ d_{22}\sin\theta = \cos\theta \\ -d_{22}\cos\theta = \sin\theta \end{cases} \quad (S12)$$

In order for the above equations to hold, di (i = 1, 2) must satisfy the following conditions:

$$\begin{cases} d_1 = -\tan\theta = -\cot\theta \\ d_2 = \tan\theta = \cot\theta \end{cases} \quad (S13)$$

Then we get:

$$\theta = 45° + n \cdot 90°, \quad n \in \mathbb{Z} \quad (S14)$$

Let $\theta=-45°$, we get $\psi_2 = \psi$. Thus, we get d1=1 and d2 = -1, with the Jones matrix in the following form:

$$J_2 = R(\psi)\begin{pmatrix} 1 & 0 \\ 0 & -1 \end{pmatrix}R^{-1}(\psi) \quad (S15)$$

Thus, the Jones matrix that accounts for the difference conversion between an orthogonal polarization pair can be written as:

$$J = J_1 + J_2 = \frac{n_1}{N}R(\psi - 45°)\begin{pmatrix} e^{-2i\chi} & 0 \\ 0 & e^{2i\chi} \end{pmatrix}R(45° - \psi) + \frac{n_2}{N}R(\psi)\begin{pmatrix} 1 & 0 \\ 0 & -1 \end{pmatrix}R(-\psi) \quad (S16)$$

**Note III: The necessity of introducing Disorder compared to Periodic construction**

Although it may seem intuitive to use a periodic supercell construction to achieve the Jones matrix presented in Equation 9 of the main text, several potential issues arise when applying periodic designs to this task. Consider the target polarization conversion parameters stated in the main text, with 2 psi=300, 2 xi = 45, and DoP = 0.65. The dimensions of the two meta-atoms involved are 320 nm by 650 nm and 600 nm by 250 nm, respectively. One might assume that arranging these meta-atoms uniformly in a supercell, as illustrated in the inset of the Fig. S2(a), would yield the desired result. An 8:3 ratio of meta-atoms is used to approximate the target ratio of 2.743, which is required for the desired DoP control. However, due to the periodic boundary conditions, when the supercell is translated, the meta-atom at the bottom-left corner (meta-atom B) aligns with the meta-atom at the top-right corner of the adjacent supercell. This leads to an excess of meta-atoms B in certain local regions, potentially diminishing the polarization conversion efficiency. This perturbation effect becomes more pronounced when the number of meta-atoms within the supercell is relatively small. In contrast, if the entire disordered metasurface itself were translationally symmetric, the impact of this perturbation would be negligible. Our simulations for this pattern yield a DoP of 0.714, with the azimuthal and elevation angles of the dominant polarization state at -66.31° and 26.56°, respectively. When evaluated using the Euclidean distance between Stokes parameters, the similarity to the target result is 0.76, which falls significantly short of the performance achieved using the disordered metasurface design described in the main text.

An alternative approach might be to arrange the meta-atoms within a two-dimensional crystal lattice, such as a rectangular or hexagonal lattice. This can be done by treating a small group of meta-atoms as a unit cell and placing multiple unit cells at lattice vertices. We provide two such examples in Figure S2(b) and (c). The design in Figure S2(b) encounters the same issues as in Figure S2(a), while the design in Figure S2(c) generates a quasi-gradient phase along the indicated direction. The white dashed rectangular line represents the larger lattice, while the circular one corresponds to the smaller meta-atom cluster in the Figure S2(b). The stokes similarity of this design is 0.82. Although the periodic arrangement in the inset of Figure S2(c) more closely resembles a typical lattice distribution, the strong internal order within the supercell is further enhanced by the periodicity. The arrow in the inset indicates a quasi-phase gradient direction, leading to beam deflection, as confirmed by the far-field intensity shown in Figure S2(c). The computed similarity of the Stokes parameters for this design is only 0.8. Additionally, the periodic structure leads to pronounced higher-order diffraction effects, as the diffraction fields in the Figure S2 shown.

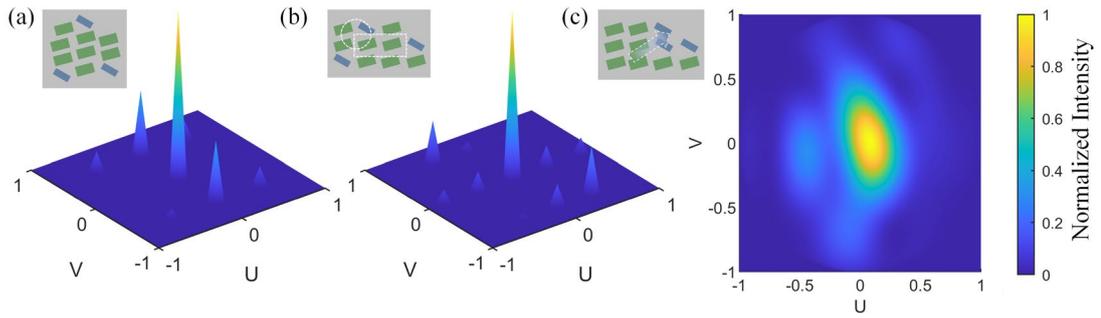

Figure S2. (a-b) Diffraction field of periodic design. (c) Intensity distribution of the far-field. Inset: metasurface arrangements for corresponding design. $U$ and $V$ are the direction cosines.

It is important to note that this discussion does not prove that a disordered metasurface is the only viable solution for achieving this target. Advanced algorithms for periodic design could be developed to address these issues and meet the desired goals. However, directly adopting a disordered design appears to be a more practical approach. Moreover, an appropriate arrangement algorithm (discussed in Note V) is crucial, as it is mathematically impossible to achieve uniform tiling of rectangles with arbitrary proportions and sizes in two dimensions using only simple translations and rotations.

**Note IV: Flow chart of the algorithm**

Fitness is defined as a measure of how closely the area ratios of different rectangle types within local regions of the plane match their ideal proportions, which are based on the given quantities of each rectangle type. At each step, fitness is evaluated by comparing the actual area distribution in the current region to the ideal, with the goal of minimizing this deviation.

The greedy nature of the algorithm lies in its selection process: at each step, only the placement that provides the maximum local improvement in fitness is chosen, without considering the global impact of this decision. The algorithm focuses on immediate, local uniformity while ignoring potential long-term effects on the overall layout. Additionally, new rectangles are placed along the current outline to prioritize dense packing, further emphasizing the greedy, short-term optimization approach.

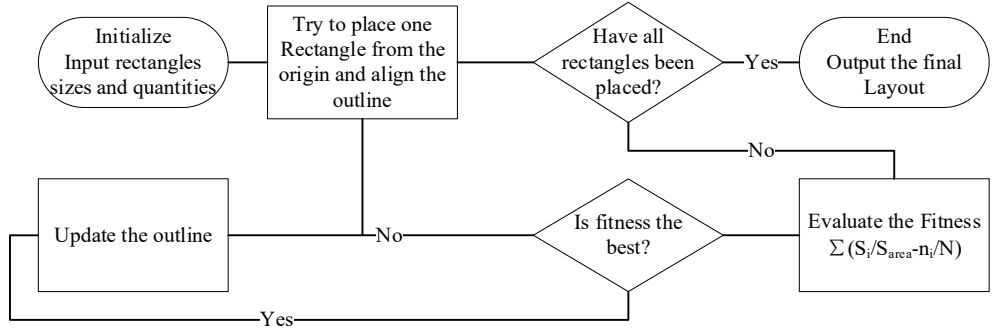

Figure S3. The flow chart of the arrangement algorithm.

**Note V: Original structural and arrangement parameters for the main text results**

We constructed a nanostructure database consisting of rectangular silicon (RI = 3.45 @ 1550 nm) nanopillars on a silica substrate (RI = 1.45 @ 1550 nm). The thickness of the nanopillars is 940 nm, and the period during parameter sweeping is set to 960 nm. The length and width are varied between 200 and 800 nm, as illustrated in Figure S4. The meta-atoms used to demonstrate the evolution along the latitude on the Poincaré sphere have dimensions of 320 nm by 650 nm and 600 nm by 250 nm. The arrangement data for $2\psi=300°$ is provided in the table S1 and S2. The effective sizes of these meta-atoms are 850 nm by 650 nm and 760 nm by 850 nm, respectively. Table TS1 lists the position coordinates of meta-atoms A, while Table TS2 provides the coordinates for meta-atoms B. For the evolution along the longitude on the Poincaré sphere, the data includes the size parameters for each set of meta-atom combinations at each longitude, which are listed in Table S3. For DoP modulation, the meta-atoms have dimensions of 220 nm by 590 nm and 690 nm by 250 nm, respectively.

The other raw data that support the findings of this study are available from the corresponding author upon reasonable request.

Table S1. Position coordinates of meta-atom with dimensions (320, 650) nm

| X | Y | X | Y | X | Y | X | Y | X | Y |
|---|---|---|---|---|---|---|---|---|---|
| 425 | 325 | 1275 | 325 | 1185 | 975 | 425 | 1825 | 1275 | 1625 |
| 2125 | 1175 | 425 | 2475 | 2275 | 1825 | 3125 | 325 | 575 | 3125 |
| 2065 | 2475 | 3125 | 975 | 1425 | 3125 | 2915 | 2475 | 2275 | 3125 |
| 1195 | 3775 | 3975 | 1175 | 3885 | 1825 | 3775 | 2475 | 1195 | 4425 |
| 4825 | 325 | 2805 | 3975 | 4825 | 975 | 2045 | 4625 | 4735 | 1825 |
| 4005 | 3125 | 425 | 5075 | 3655 | 3975 | 1275 | 5275 | 2895 | 4625 |
| 5675 | 325 | 4505 | 3775 | 5615 | 1825 | 2885 | 5275 | 1185 | 5925 |
| 5615 | 2475 | 5355 | 3325 | 4505 | 4425 | 2035 | 6125 | 6525 | 325 |
| *425*  | *6575* | *2885* | *5925* | *6525* | *975*  | *3735* | *5475* | *6465* | *1625* |
| *4585* | *5075* | *6465* | *2275* | *1275* | *6775* | *6205* | *3125* | *2885* | *6575* |
| *425*  | *7225* | *5445* | *4825* | *4585* | *5725* | *6295* | *3775* | *7315* | *1625* |
| *6315* | *4425* | *1335* | *7625* | *2885* | *7225* | *4495* | *6375* | *7165* | *3125* |
| *3735* | *7025* | *6295* | *5075* | *8165* | *325*  | *2185* | *7875* | *8165* | *975*  |
| *5345* | *6375* | *8165* | *1625* | *1205* | *8275* | *4585* | *7025* | *8095* | *2275* |
| *3035* | *7875* | *6215* | *5725* | *7195* | *4625* | *8095* | *2925* | *2055* | *8525* |
| *8045* | *3575* | *6215* | *6375* | *425*  | *8925* | *9035* | *1175* | *8085* | *4225* |
| *4655* | *7875* | *3415* | *8525* | *8945* | *2675* | *7065* | *6125* | *6195* | *7025* |
| *8045* | *4875* | *2115* | *9175* | *4265* | *8525* | *425*  | *9575* | *5505* | *7875* |
| *2965* | *9175* | *7915* | *5525* | *8935* | *4175* | *9885* | *325*  | *1275* | *9825* |

Table S2. Position coordinates of meta-atom with dimensions (600, 250) nm.

| X | Y | X | Y | X | Y | X | Y | X | Y |
|------|------|------|------|------|------|------|------|------|------|
| 380  | 1075 | 2080 | 425  | 1230 | 2375 | 3080 | 1725 | 390  | 3875 |
| 3930 | 425  | 2000 | 3875 | 3200 | 3225 | 4810 | 2575 | 5630 | 1075 |
| 2080 | 5375 | 380  | 5825 | 3700 | 4725 | 5310 | 4075 | 2080 | 6875 |
| 3690 | 6225 | 7330 | 425  | 7290 | 2375 | 5410 | 5575 | 400  | 7975 |
| 7120 | 3875 | 3850 | 7775 | 7100 | 5375 | 5390 | 7125 | 9030 | 425  |
| 1250 | 9025 | 8970 | 1925 | 8900 | 3425 | 7020 | 6875 | 5100 | 8625 |
| 8890 | 4925 | 2380 | 9925 | 380  | 10325| 10400| 1075 | 8470 | 6275 |

Table S3. Structure parameters for demonstrating polarization evolution along a longitude on the Poincaré sphere.

| 2χ | -90 | -67.5 | -45 | -22.5 | 0 | 22.5 | 45 | 67.5 | 90 |
|---|---|---|---|---|---|---|---|---|---|
| Meta-atom A (L,W) nm | 470,250 | 240,510 | 340,750 | 560,370 | 550,550 | 360,520 | 320,650 | 240,500 | 250,470 |
| Meta-atom B (L,W) nm | 250,680 | 680,250 | 250,760 | 610,250 | 690,250 | 580,250 | 600,250 | 250,690 | 250,680 |

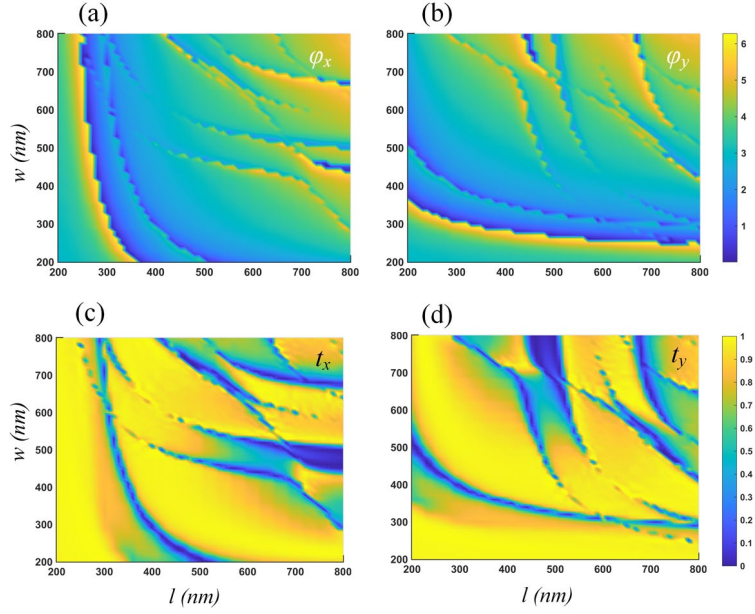

Figure S4 (a-b) Phase delay and (c-d) transmission coefficients with respect to the meta-atoms' length (l) and width (w) at a wavelength of 1550 nm for x-polarized and y-polarized incident light.

**Note VI: Forward Method for Implementing Arbitrary Jones Matrices**

In addition to the modulation described in the main text, the proposed homogeneous disordered metasurface can be applied to realize arbitrary Jones matrices when $M>2$. While many previous works [1-6] have demonstrated that a diatomic design is sufficient to construct arbitrary Jones matrices for single-layer surfaces, the approach of providing more degrees of freedom (DoFs) offers greater flexibility in the selection of meta-atoms, as well as in other design aspects. For an arbitrary Jones matrix, it can be expressed in the following form:

$$J = \begin{bmatrix} J_{11}e^{i\varphi_{11}} & J_{12}e^{i\varphi_{12}} \\ J_{21}e^{i\varphi_{21}} & J_{22}e^{i\varphi_{22}} \end{bmatrix} = \begin{pmatrix} t_{xx'} & t_{yx'} \\ t_{xy'} & t_{yy'} \end{pmatrix} \quad (S17)$$

The off-diagonal elements of the matrix are symmetric due to mirror symmetry, which is characteristic of single-layer structures [2]. By applying a 45-degree rotation, the off-diagonal terms can be rotated into the following configuration:

$$\begin{aligned} J &= \begin{bmatrix} J_{11}e^{i\varphi_{11}} & J_{12}e^{i\varphi_{12}} \\ J_{21}e^{i\varphi_{21}} & J_{22}e^{i\varphi_{22}} \end{bmatrix} \\ &= \begin{bmatrix} J_{11}e^{i\varphi_{11}} & 0 \\ 0 & J_{22}e^{i\varphi_{22}} \end{bmatrix} + R(45)\begin{bmatrix} J_{12}e^{i\varphi_{12}} & 0 \\ 0 & -J_{12}e^{i\varphi_{12}} \end{bmatrix} R(-45) \quad (S18) \\ &= \begin{bmatrix} J_{11}e^{i\varphi_{11}} & 0 \\ 0 & J_{22}e^{i\varphi_{22}} \end{bmatrix} + J_{12}R(45)\begin{bmatrix} e^{i\varphi_{12}} & 0 \\ 0 & e^{i(\varphi_{12}+\pi)} \end{bmatrix} R(-45) \end{aligned}$$

The second term can be interpreted as the Jones matrix of a birefringent meta-atom, with fast and slow axes having phase delays of $\phi_{12}$ and $\phi_{12}+\pi$, respectively. The amplitude of $J_{12}$ can be retrieved after determining the quantity ratio from the first term. Since the second term has the same form as the Jones matrix of a meta-atom, we only need to decompose the first term. To solve for this decomposition, we can construct the following equation:

$$\begin{bmatrix} J_{11}e^{i\varphi_{11}} & 0 \\ 0 & J_{22}e^{i\varphi_{22}} \end{bmatrix} = \sum_{i=1}^{N} \frac{1}{S_i} J_{atom}^{i}$$
$$= \begin{bmatrix} xe^{i\varphi_{x1}} & 0 \\ 0 & xe^{i\varphi_{x2}} \end{bmatrix} + \begin{bmatrix} ye^{i\varphi_{y1}} & 0 \\ 0 & ye^{i\varphi_{y1}} \end{bmatrix} \quad (S19)$$

And we can obtain the following system of equations:

$$\begin{cases} J_{11}e^{i\varphi_{11}} = xe^{i\varphi_{x1}} + ye^{i\varphi_{y1}} \\ J_{22}e^{i\varphi_{22}} = xe^{i\varphi_{x2}} + ye^{i\varphi_{y1}} \end{cases} \quad (S20)$$

We can observe that there are four equations with six variables, indicating that the system has an infinite number of solutions. In this case, it is possible to impose initial conditions on the system to obtain a desired solution. For example, by setting $\phi_{11}=\phi_{x1}=\phi_{y1}$, we get $J_{11} = x+y$. The second equation can then be solved analytically, which can be easily visualized using an Argand diagram shown in Figure S5 (a).

As an example, consider constructing a Jones matrix that converts 45° linearly polarized (LP) light to right circularly polarized (RCP) light. The required Jones matrix for this conversion is:

$$J = \begin{pmatrix} 1.2645 - 0.1285i & -0.2645 + 0.1285i \\ -0.2645 - 0.1285i & 0.2645 + 0.8715i \end{pmatrix} \quad (S21)$$

Let $x=y$ and $\phi_{11}=\phi_{x1}=\phi_{y1}$. We can then solve for the solution as follows:

$$\begin{cases} x = y = J_{11}/2 = 0.6355 \\ \varphi_{x1} = \varphi_{y1} = \varphi_{11} = -0.1013 \\ \varphi_{x2} = \varphi_{22} - \arccos(J_{22}/J_{11}) = 2.048 \\ \varphi_{y2} = 2\varphi_{22} - \varphi_{x2} = 0.5041 \end{cases} \quad (S22)$$

The corresponding meta-atoms have been selected with dimensions of (730, 240), (220, 290), and (250, 650) nm, respectively, with a quantity ratio of 13:13:6. A portion of the arrangement results is shown in Figure S5 (b). To evaluate the accuracy of the polarization conversion, we employ the fidelity, defined as the inner product between the simulated Jones vector and the target Jones vector. The fidelity achieved in this case exceeds 98%.

This example demonstrates that the proposed metasurface and its corresponding construction algorithm can be extended to various configurations of meta-atom combinations. The flexibility in design allows for greater freedom in selecting meta-atoms, and initial conditions can be adjusted according to specific requirements. This design method, with its increased degrees of freedom, allows for a direct mapping between the metasurface's physical parameters and the desired control targets. As a result, the metasurface can be flexibly designed to achieve precise light field manipulation tailored to specific application requirements.

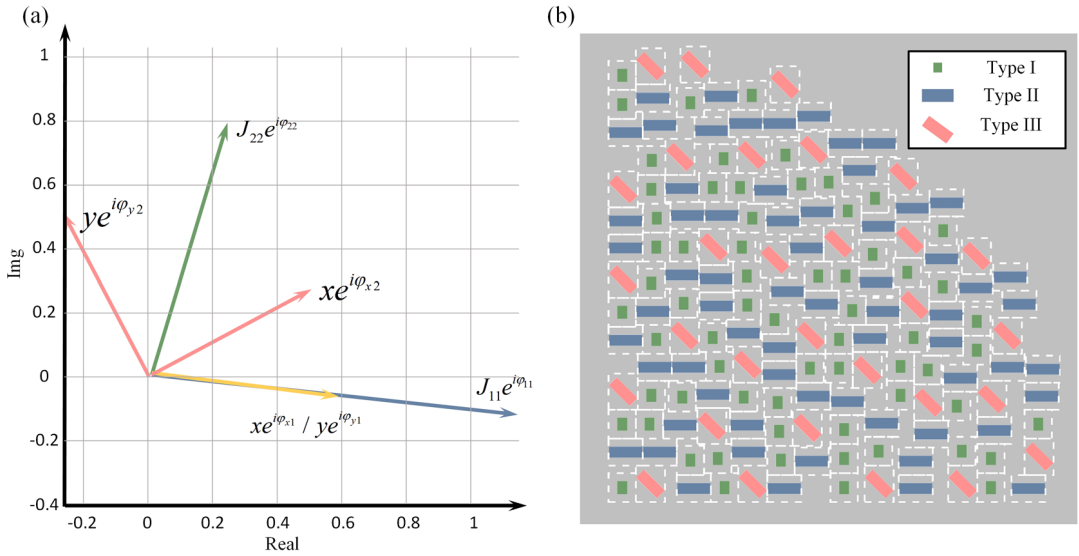

Figure S5. (a) Argand diagram to solve the complex equation (Equation S20). (b) Metasurface arrangement of the demonstration for the construction of an arbitrary Jones matrix.